\documentclass[useAMS,usenatbib]{mn2e}
\usepackage[pdftex]{graphicx}
\usepackage{hyperref}
\usepackage{amssymb}
\usepackage{epstopdf}
\usepackage[all]{hypcap}

\title[Geometrically thin accretion disk around Maclaurin spheroid]{Geometrically thin accretion disk around Maclaurin spheroid}
\author[B. Mishra, B. Vaidya]{B. Mishra$^{1}$\thanks{E-mail:
mbhupe@camk.edu.pl}, B. Vaidya$^{2}$\\
$^{1}$Copernicus Astronomical Center, Bartycka 18, Warsaw,00-716, Poland\\
$^{2}$Dipartimento di Fisica `Amedeo Avogadro' Università degli Studi di Torino, Via Pietro Giuria 1, 10125 Torino, Italy}
\begin{document}
\date{Accepted *** Received ***}
\pagerange{\pageref{firstpage}--\pageref{lastpage}} \pubyear{2014}
\maketitle
\label{firstpage}
\begin{abstract}
We investigated a semi-analytic and numerical model to study the geometrically thin and optically thick accretion disk around Maclaurin spheroid (MS). The main interest is in the inner region of the so called $\alpha$-disk, $\alpha$ being the viscosity parameter. Analytical calculations are done assuming radiation pressure and gas pressure dominated for close to Eddington mass accretion rate and $\dot{M} \lesssim 0.1\dot{M}_{Edd}$ respectively. We found that the change in eccentricity of MS gives a change at high frequency region in the emitted spectra. We found that disk parameters are dependent on eccentricity of MS. Our semi-analytic results show that qualitatively an increase in eccentricity of MS has same behavior as decrease in mass accretion rate. Numerical work has been carried out to see the viscous time evolution of the accretion disk around MS. In numerical model we showed that if the eccentricity of the object is high the matter will diffuse slowly during its viscous evolution. This gives a clue that how spin-up or spin-down can change the time evolution of the accretion disk using a simple Newtonian approach. The change in spectra can be used to determine the eccentricity of MS and thus period of the MS.
\end{abstract}
\section{Introduction}
Accretion disk remains a challenging problem mainly in the context of its viscosity prescription and radiative processes. The best known models for describing the accretion disk were proposed by \citet{1973A&A....24..337S}, \citet{1973blho.conf..343N} and \citet{1974MNRAS.168..603L} also see \citet{2013LRR....16....1A} for review. The underling assumption behind all these models was a geometrically thin accretion disk and because of this assumption accreting gas is cool as compared to the local virial temperature. There was also an assumption of equilibrium between viscous energy generation inside the disk and radiative cooling from the surface of the disk, where radiative cooling was computed with an assumption that disk is optically thick in the vertical direction. \citet{1973A&A....24..337S} applied this model to study accretion disk around non-rotating black hole assuming spherically symmetric potential. Due to its general applicability this model has also been extended to study accretion disk around stars \citep{2009ApJ...702..567V, 2009ApJ...705.1206C}, planets \citep{2014ApJ...790...78F, 2014MNRAS.440...89K}, cataclysmic variables \citep{1991A&A...249..574L, 2002MmSAI..73..206S}.

\citet{1973blho.conf..343N} extended \citet{1973A&A....24..337S} model to study thin accretion disk around rotating black hole using Kerr space time. The accretion disk around quark-stars has been studied using relativistic approach in \citet{2009A&A...500..621K}. \citet{2000ApJ...542..473B} studied accretion disks around rapidly rotating neutron stars taking into account full effects of general relativity. In which authors also studied the general relativistic spectra from the disks around rotating neutron stars with different equation of states, spin and mass. Recently \citet{2014PhRvD..89j4001G} also studied the effect of oblateness of quark-star on the orbital frequencies to investigate the QPOs. Quark stars have also been in interest to compare the properties of star with observations. \citet{2001ESASP.459..301K} studied the range of rotational frequencies and masses of strange stars and compared them with observations.

\citet{1993A&A...274..796B} studied that how mass accretion can change the eccentricity of the rapidly rotating star. Authors also suggested that this behavior can be applicable in case of neutron stars, where a change in eccentricity can cause luminosity fluctuations. Recently \citet{2014} used pseudo Newtonian potential described in \citet{2002MNRAS.335L..29K} and MS potential to investigate eigenmodes of trapped horizontal oscillations in accretion disk. We performed first time a semi-analytical and numerical study of viscous accretion disk around an accreting source which has spheroidal shape rather than spherical. We used MS potential \citep{1969efe..book.....C} for a constant density and mass of the central object. We investigated that how the non-keplerian angular velocity will change the dynamics of the accreting matter into the MS. Our goal is to study mainly inner region of the accretion disk where the multipole effects of chosen MS potential dominates. We followed the same model as \citet{1973A&A....24..337S} just by changing the potential of the central object for that of MS. We used same constant $\alpha$ viscosity prescription to proceed analytic and numerical work. We did the analytical calculation of various parameters of thin disk and we shall focus only in the inner region which may be either radiation or gas pressure dominated depending on mass accretion rate. For close to Eddington mass accretion rate disk will be radiation pressure dominated in the very inner regions and for mass accretion rate, $\dot{M} \lesssim 0.1\dot{M}_{Edd}$ it will be gas pressure dominated \citep{1973A&A....24..337S}.

We took into account the main property of MS potential which corresponds to innermost stable circular orbit (ISCO) even in Newtonian dynamics \citep{2002A&A...381L..21A,2013MNRAS.434.2825K}. We chose a constant density and mass MS and assumed that it is rotating rapidly. During its rapid rotation it can change its eccentricity and so semi-major axis. The eccentricity decreases or increases, depending on either the semi-major axis decreases or increases. In \citet{2013MNRAS.434.2825K}, it has been shown that if the eccentricity is less than a critical value of $e_c = 0.8345$, the ISCO will lie on the equator of the accreting source but if it is higher than this critical value it will be detached from the surface of the star. Keeping this change in mind we investigated the cases where eccentricity is less than critical limit. We see a change in inner radius of the accretion disk with change in eccentricity because the semi-major axis of the accreting MS is changing. This change in the inner radius of the accretion disk due to change in eccentricity will clearly change the calculated parameters like surface density, height and emitted spectra of the accretion disk. We studied the time evolution of the accretion disk around MS by solving the diffusion equation for the accreting matter. We again assumed the MS potential to proceed with the study of non-stationary disk. In numerical work we studied how change in eccentricity will change the time evolution of the accretion disk. 

The article is organized in the following manner. In \S 2 we describe physical model of accretion disk which covers steady thin disk and also numerical study of the time evolution of the accretion disk. \S3 is devoted for describing all the results we obtained analytically and numerically. In \S4 we discuss all the results described in \S3 and we conclude in \S5 with future applications of our accretion disk model around MS.  
\begin{figure}
\centering
\includegraphics[width=1\columnwidth]{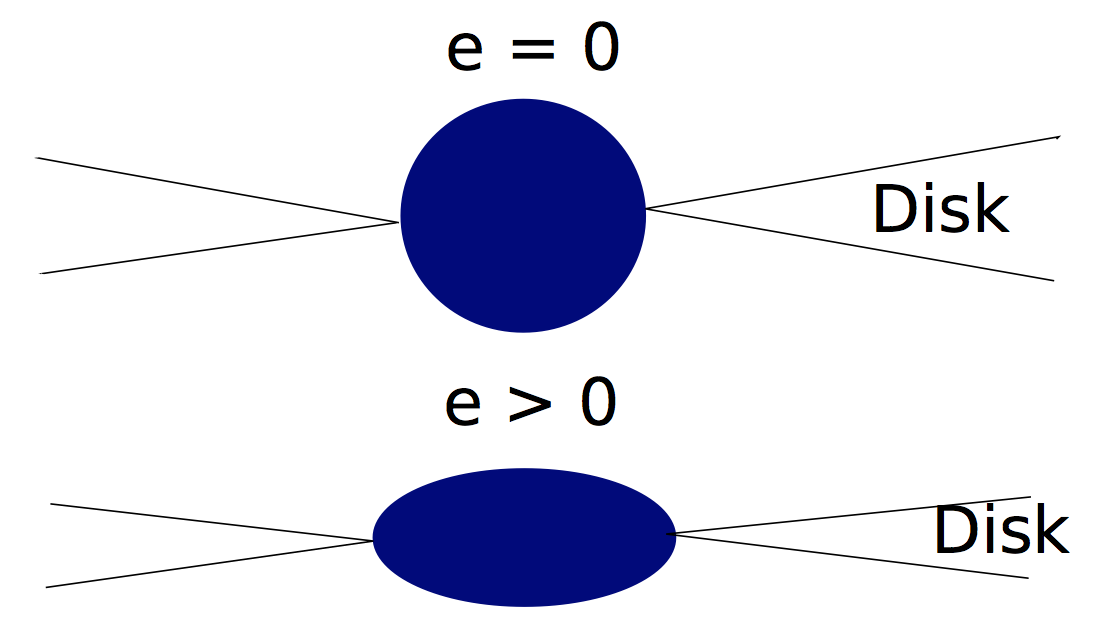}
\caption{\small{Sketch of the accretion disk around MS. For higher value of eccentricity $e$ the MS is more oblate and hence inner radius of the disk is larger}.}
\label{fig1}
\end{figure}
\section{Physical Model}
\subsection{Maclaurian Spheroid}
In our model the semi-major axis, $a(e)$ of the MS changes with eccentricity. We also assumed in our model that the disk inner radius is same as semi-major axis of the MS, i.e. disk terminates at the equator of the MS. This assumption causes a change in inner radius of the accretion disk due to change in eccentricity.
\begin{equation}
a(e) = \frac{R_0}{(1 - e^2)^{1/6}}
\label{semi}
\end{equation}
where $a(e)$ is the semi-major axis of the MS and $R_0 = a(0)$ is the radius of spherical MS. The maximum value of eccentricity in this paper is $e = 0.8345$. The reason of this limit is due to fact that in case of potential for MS the maximum of radial epicyclic frequency lies at $r = \sqrt{2}ae$ for spheroid eccentricities $e>1/\sqrt{2}$ but it vanishes for $e_c = 0.83458318$ at the equator of the star \citep{2013MNRAS.434.2825K}. If we still increase the eccentricity further $e>e_c$, the innermost stable circular orbit (ISCO) will be separated from the equator of star and it will be at $r_{ms} = 1.198203ae$ \citep{2013MNRAS.434.2825K}. So we always kept the inner radius at no-torque boundary (the radius at which there is zero viscosity) which coincides with the variable semi-major axis of MS. One can increase the eccentricity further to investigate the accretion disk for which the inner radius does not lie at the surface of MS but in this paper we shall not discuss it. The angular velocity in case of MS potential is given by
\begin{equation}
\Omega ^2 \left(e,r\right)= 2\pi G\rho_* (1-e^2)^{1/2}e^{-3}\left[\gamma_r - \cos \gamma_r \sin\gamma_r \right]
\label{omega}
\end{equation}  
where $\gamma_r = \arcsin (a e/r)$, $a$ is the semi-major axis of MS and $\rho_*$ is density of the MS \citep{2013MNRAS.434.2825K}. Now we have angular velocity of the matter for chosen potential, next goal is to follow \citet{1973A&A....24..337S} disk model and do the calculations for angular velocity calculated from Eq.~\ref{omega}. This analytic approach gave us various disk parameters like, half-thickness, surface density, disk central temperature and radial velocity in the inner region of the accretion disk.
\subsection{Steady thin accretion disk}
We considered a geometrically thin accretion disk (height of the accretion disk is much smaller than its radial structure) around MS (Fig.\ref{fig1}). Calculations are done in cylindrical coordinate system $(r,\phi,z)$, assuming azimuthal symmetry. The main interest of this model is to study the steady-state disk to see the behavior of disk parameters and emitted spectra form the accretion disk. To proceed the calculations for steady-state disk we used Eq.~\ref{omega} and Eq.~\ref{angmom} to analytically calculate disk parameters in the inner region of the accretion disk. In this section we shall present the calculations for radiation pressure dominated as well as gas pressure dominated region (opacity due to electron scattering). The disk parameters for gas pressure dominated region with opacity due to free-free absorption is presented in appendix~\ref{a1}. The angular momentum equation in terms of angular velocity is given by
\begin{equation}
\frac{\Sigma d\Omega r^2}{dt} = -\Sigma v_r\frac{d\Omega r^2}{dr} = \frac{1}{r}\frac{d}{dr}W_{r\phi}r^2,
\label{angmom}
\end{equation}
where $W_{r\phi}$ is the stress between adjacent layers \citep{1973A&A....24..337S}, which is assumed to be a function of sound speed, $v_s$ and surface density $\Sigma$.
\begin{equation}
\Sigma = 2\int_0^{z_0} \rho dz,
\end{equation}
\begin{equation}
W_{r\phi} = -\alpha\Sigma v_s^2,
\end{equation}
In stationary disk model $\dot{M} = 2\pi\Sigma v_r r =$ const and $v_r < 0$. Now integrating Eq.~\ref{angmom} we obtain
\begin{equation}
\dot{M}\Omega r^2 = -2\pi W_{r\phi}r^2 + C,
\label{const}
\end{equation}
where $C$ is constant which we calculated by using no-torque boundary condition \citep{1973A&A....24..337S}. Finally we get the equation to calculate disk parameters in all three regions of the accretion disk.
\begin{equation}
\dot{M}\left(\Omega r^2 - \Omega(a)a^2\right) = 2\pi\alpha\Sigma v_s^2 r^2.
\label{keyeqn}
\end{equation}
Now the energy flux radiated from the surface unit as function of $\Omega (e,r)$ is given by
\begin{equation}
Q = -\frac{\dot{M}\left(\Omega r^2 - \Omega(a)a^2\right)}{4\pi r}\frac{d\Omega}{dr}.
\label{radflux}
\end{equation}
using Eq.~\ref{keyeqn} and Eq.~\ref{radflux} together with assumption of radiation and gas pressure dominated disk we calculated disk thickness, surface density, temperature and radial velocity in the inner region of the accretion disk.
\subsubsection{Radiation pressure dominated region}
In steady $\alpha$-disk the total pressure is due to contribution from gas pressure $P_{gas}$ and radiation pressure $P_{rad}$. If the mass accretion rate is close to its Eddington value the inner region of disk is radiation pressure dominated where in the interaction of matter and radiation electron scattering on free electrons has dominating contribution. We substituted $\Omega(e,r)$ in Eq.~\ref{keyeqn} and Eq.~\ref{radflux} from Eq.~\ref{omega} to calculate the disk parameters like disk half thickness $z_0(r)$, surface density $\Sigma(r)$, disk central temperature $T(r)$ and radial velocity $v_r(r)$ of the matter. We expressed the analytic expression in terms of defined parameters $\gamma_r$, $\gamma_a$, $p_r$, $p_a$ and $k_1$ to simplify the long expressions.
\begin{equation}
z_0(r) = \frac{\sigma_T\dot{M}\sin^2\gamma_r\tan\gamma_r\left(1 - \left(\frac{p_a}{p_r}\right)^{1/2}\left(\frac{a}{r}\right)^2\right)}{8\pi p_r c},
\label{z}
\end{equation}
\begin{equation}
\Sigma(r) = \frac{32\pi c^2 p_r^{1/2}\left(1 - \left(\frac{p_a}{p_r}\right)^{1/2}\left(\frac{a}{r}\right)^2\right)^{-1}}{\alpha\sigma_T^2 k_1^{1/2}\tan^2{\gamma_r}\sin{\gamma_r}\dot{M}},
\label{sigma}
\end{equation}
\begin{equation}
\varepsilon(r) = \frac{6c p_r^{3/2}k_1^{1/2}}{\alpha\sigma_T\sin^2\gamma_r\tan\gamma_r},
\end{equation}
\begin{equation}
T(r) = \left(\frac{\varepsilon(r)}{b}\right)^{1/4},
\label{midplane}
\end{equation}
\begin{equation}
\tau(r) = \sqrt{0.11\sigma_T T^{-7/2}n}\Sigma,
\end{equation}
\begin{equation}
n(r) = \frac{\Sigma}{2m_pz_0},
\label{nr}
\end{equation}
\begin{equation}
v_r(r) = \frac{\dot{M}}{2\pi\Sigma r},
\label{vel}
\end{equation}
where,
\begin{equation}
k_1 = 2\pi G\rho_* \frac{(1-e^2)^{1/2}}{e^3},
\label{k1}
\end{equation}
\begin{equation} 
\gamma_r = \arcsin\left(\frac{ae}{r}\right), 
\label{gaammar}
\end{equation}
\begin{equation}
\gamma_a = \arcsin(e),
\end{equation}
\begin{equation}
p_r = (\gamma_r - \sin\gamma_r\cos\gamma_r),
\label{pr}
\end{equation}
\begin{equation}
p_a = (\gamma_a - \sin\gamma_a\cos\gamma_a).
\label{pa}
\end{equation}
$\dot{M}$ is mass accretion rate, $\sigma_T$ is mean opacity of the matter for Thomson electron scattering, $\alpha = 0.01$ is viscosity coefficient, $m_p$ is the mass of proton, $b = 3\sigma_b/c$ where $\sigma_b$ is Stefan-Boltzmann constant, $\varepsilon(r)$ is radial distribution of energy density, $\tau(r)$ is optical depth, $n(r)$ is the number density of the steady thin accretion disk.

We also confirmed that region we considered is radiation pressure dominated for mass accretion rate close to Eddington mass accretion rate, $\dot{M} = 0.8\dot{M}_{Edd}$. This mass accretion rate makes the disk radiation pressure dominated for very inner regions of the disk. In this region ratio of radiation pressure to gas pressure is larger than unity, where gas pressure is calculated using ideal gas equation of state. The Eddington mass accretion rate for the MS we considered is $\dot{M}_{Edd} \approx 10^{17}\textrm{g}$ $\textrm{s}^{-1}$ for MS mass $M_* = 2.1M_{\odot}$. We also see from Eq.~\ref{z} and Eq.~\ref{nr} which have dependence on mass accretion rate $\dot{M}$ which makes $P_{gas}$ to depend on mass accretion rate $\dot{M}^{-2}$. This corresponds that if we increase the mass accretion rate the gas pressure $P_{gas}$ will decrease further. This will maintain our assumption of radiation pressure dominated region in the inner region of the accretion disk.
\subsubsection{Gas pressure dominated region, electron scattering}
In case of mass accretion rate very low as compare to Eddington mass accretion rate, $\dot{M} \lesssim 0.1\dot{M}_{Edd}$ the disk will be gas pressure dominated $(P_g >> P_r)$. We assumed that in this case electron scattering gives main contribution to opacity. The sound speed in this region is given by $v^2_s = kT/m_p$. The disk parameters are governed by following power laws,
\begin{equation}
\Sigma(r) = \left(\frac{bck_1^{3/2}\dot{M}^3m_p^4p_r^{3/2}}{3\sigma_T\pi^3\alpha^4k^4r}\right)^{\frac{1}{5}}\left[1 - \left(\frac{p_a}{p_r}\right)^{1/2}\left(\frac{a}{r}\right)^2\right]^{3/5},
\label{gasd}
\end{equation}
\begin{equation}
\varepsilon(r) = \frac{3Q\sigma_T\Sigma(r)}{4c},
\end{equation}
\begin{equation}
T(r) = \left(\frac{\varepsilon(r)}{b}\right)^{1/4},
\end{equation}
\begin{equation}
z_0(r) = \sqrt{\frac{k T(r)}{m_p\Omega^2}},
\end{equation}
\begin{equation}
n(r) = \frac{\Sigma(r)}{2m_p z_0(r)},
\end{equation}
\begin{equation}
\tau(r) = \sqrt{\sigma_T \sigma_{ff}}\Sigma(r),
\end{equation}
\begin{equation}
v_r = \frac{\dot{M}}{2\pi\Sigma(r) r}.
\label{gasv}
\end{equation}
where, $\Omega$ is orbital frequency given by Eq.\ref{omega}, $\Sigma(r)$ is the height integrated surface density, $T(r)$ is the disk central temperature, $z_0(r)$ is the half thickness of the disk, $n(r)$ is the number density, $\tau(r)$ is the optical depth and $v_r$ is the radial velocity. $\sigma_T$ is opacity, $c$ is speed of light, $\alpha$ is angular momentum transportation factor, $m_p$ is the mass of proton and $\sigma_{ff}$ is opacity due to free-free absorption.
\subsection{Non-stationary accretion disk}
In non-stationary model we numerically studied the time evolution of the thin accretion disk around MS. The viscosity mechanism causes a transport of angular momentum outwards and matter accretion inwards. We numerically integrated the diffusion equation Eq.~\ref{diffusion} with constant viscosity $\nu$ as a first approximation. We used Crank-Nicolson method which is described in \citet{2010A&A...513A..79B} to solve the linear diffusion-advection equation in code units. We wrote the diffusion equation in terms of angular velocity so these equations can be fed with either MS potential or any different potential or angular velocity to see the non-stationary behavior of the viscous accretion disk.
\begin{equation}
\frac{\partial\Sigma}{\partial t} = -\frac{1}{r}\frac{\partial}{\partial r}\left[\frac{1}{2\pi(r^2\Omega)^\prime}\frac{\partial G}{\partial r}\right]
\label{diffusion}
\end{equation}
\begin{equation}
v_r = \frac{1}{2\pi r\Sigma (r^2\Omega)^\prime}\frac{\partial G}{\partial r}
\label{velocity}
\end{equation}
where,
\begin{equation}
G(r,t) = 2\pi r\nu\Sigma r^2 \Omega^\prime 
\end{equation}
is the torque exerted by two adjacent rings to each other in the accreting matter. Now if we choose the Keplerian angular velocity the above equations reduce to diffusion equation used for study of accretion disk evolution by various models based on $\alpha$-disk model \citep{1974MNRAS.168..603L}.
\begin{equation}
\frac{\partial\Sigma}{\partial t} = \frac{3}{r}\frac{\partial}{\partial r}\left[r^{1/2}\frac{\partial}{\partial r}\left(\nu\Sigma r^{1/2}\right)\right]
\end{equation}
\begin{equation}
v_r = -\frac{3}{\Sigma r^{1/2}}\frac{\partial}{\partial r}\left(\nu\Sigma r^{1/2}\right)
\end{equation}
where $\Sigma$ is the surface density, $\nu$ is kinematic viscosity. Now using Eq.~\ref{omega}, Eq.~\ref{diffusion} and Eq.~\ref{velocity} we shall compute the time evolution of the accretion disk for different eccentricities $e$ of the MS. Depending on eccentricity $e$, matter can be diffused either rapidly or slowly. 
\section{Results}
\subsection{Steady state disk}
In this section we shall use Eq.~\ref{z} to Eq.~\ref{vel} and Eq.~\ref{gasd} to Eq.~\ref{gasv} to describe the results in inner region of the accretion disk. The radius of MS for $e = 0.0$ is $R_0 = 10^6 \textrm{cm}$. The constant density of MS, $\rho_* = 10^{15}\textrm{g}$ $\textrm{cm}^{-3}$. In this model to calculate disk parameters we changed the eccentricity, $e$ of MS to see the effect on the disk thickness, surface density, temperature and radial velocity of accreting matter in the inner region of accretion disk. We chose two extreme limits of eccentricity which are $e = 10^{-4}$, $e = 0.8345$ and an intermediate value of $e = 0.2$. 

\begin{figure*}
\centering
\includegraphics[width=2\columnwidth]{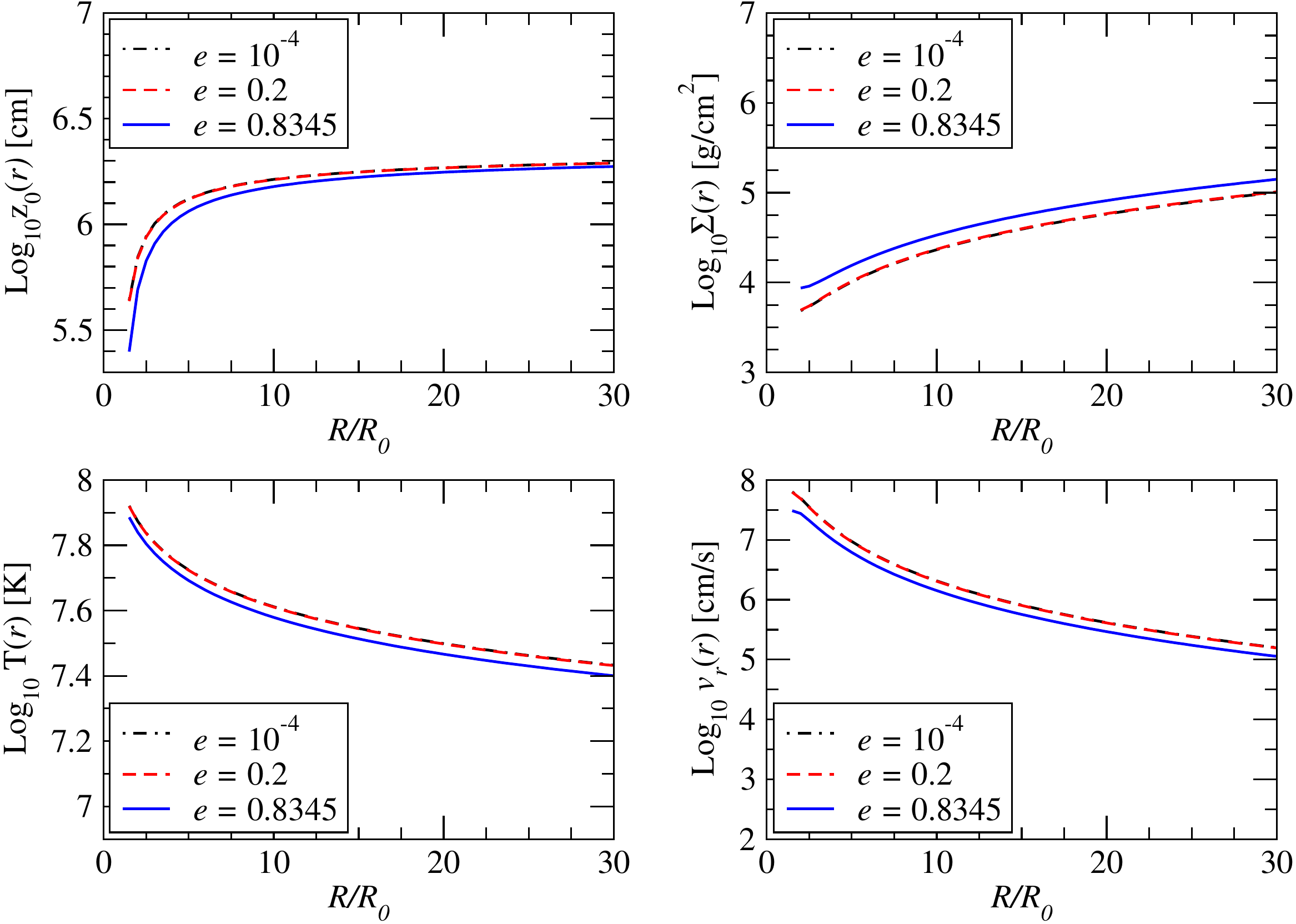}
\caption{\small{Multiplot shows logarithmic radial variation of height $z_0(r)$, surface density $\Sigma(r)$, central temperature $T(r)$ and radial velocity $v_r(r)$ for radiation pressure dominated region. The mass accretion rate for this plot is $\dot{M} = 0.8\dot{M}_{Edd}$. The different chosen eccentricities are shown by $e = 10^{-4}$ (black dotted-dashed curve), $e = 0.2$ (red dashed curve) and $e = 0.8345$ (solid blue curve). This color convention for chosen eccentricities is same throughout the article. The left upper panel shows the logarithmic radial variation of half-thickness $z_0(r)$ of the disk. The right upper panel shows the logarithmic radial variation of the surface density $\Sigma(r)$. The left lower panel shows the logarithmic radial variation of the disk central temperature $T(r)$. The lower right panel shows the logarithmic radial variation of radial velocity $v_r(r)$ of the accreting matter.}}
\label{fig2}
\end{figure*}

\begin{figure*}
\centering
\includegraphics[width=2\columnwidth]{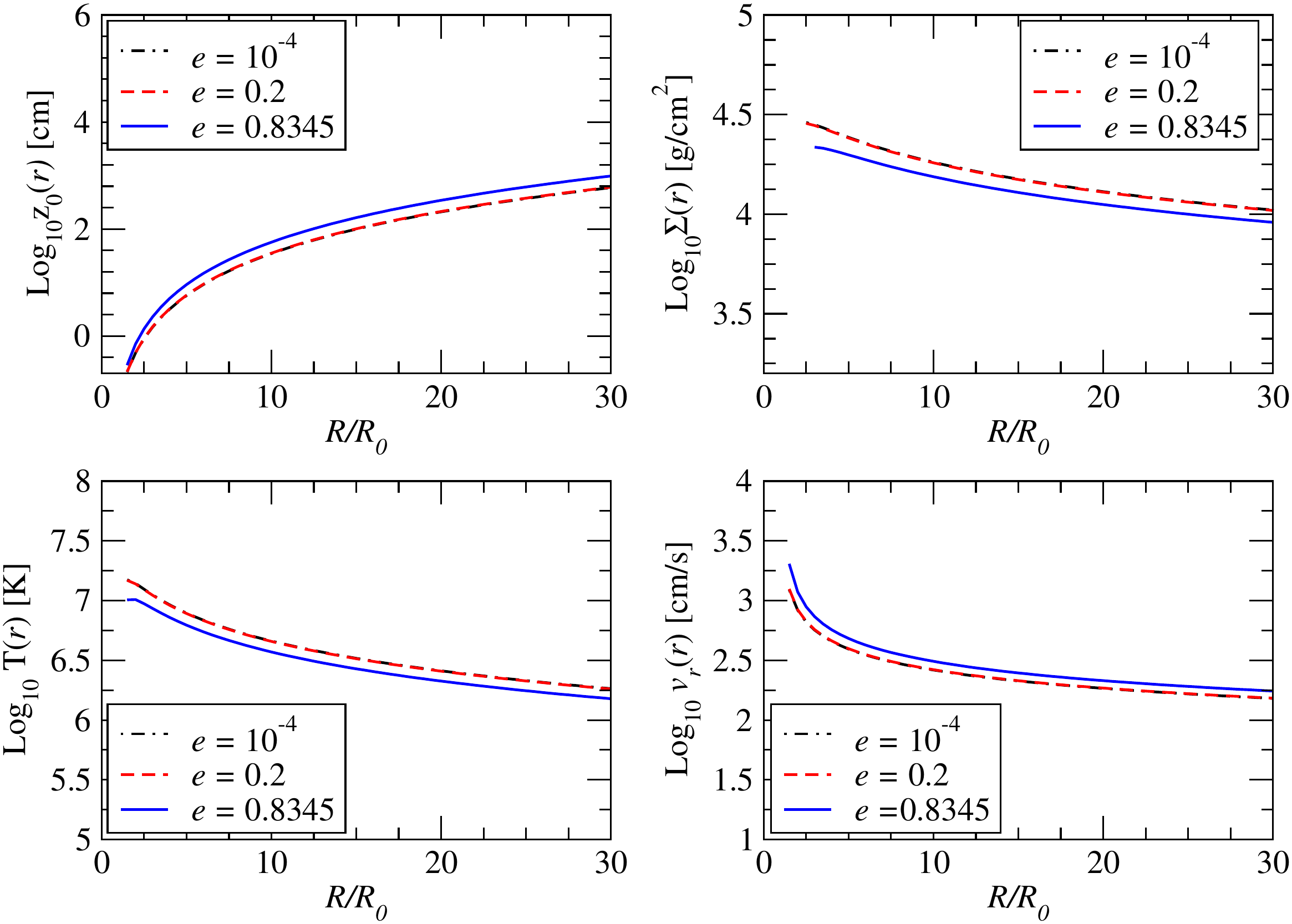}
\caption{\small{Multiplot shows logarithmic radial variation of height $z_0(r)$, surface density $\Sigma(r)$, central temperature $T(r)$ and radial velocity $v_r(r)$ for gas pressure dominated accretion disk. The mass accretion rate for this plot is $\dot{M} = 10^{-4}\dot{M}_{Edd}$ (gas pressure dominated disk). The different chosen eccentricities are shown by $e = 10^{-4}$ (black dotted-dashed curve), $e = 0.2$ (red dashed curve) and $e = 0.8345$ (solid blue curve). This color convention for chosen eccentricities is same throughout the article. The left upper panel shows the logarithmic radial variation of half-thickness $z_0(r)$ of the disk. The right upper panel shows the logarithmic radial variation of the surface density $\Sigma(r)$. The left lower panel shows the logarithmic radial variation of the disk central temperature $T(r)$. The lower right panel shows the logarithmic radial variation of radial velocity $v_r(r)$ of the accreting matter.}}
\label{fig3}
\end{figure*}

Fig.~\ref{fig2} and Fig.~\ref{fig3} show the radial variation of half-thickness $z_0(r)$, surface density $\Sigma(r)$, disk central temperature $T(r)$ and radial velocity $v_r(r)$ for $\dot{M} = 0.8\dot{M}_{Edd}$ mass accretion rate (radiation pressure dominated disk) and $\dot{M} = 10^{-4}\dot{M}_{Edd}$ (gas pressure dominated disk) respectively. In this case the inner grid point of the plot for all the parameters is $2a$ to avoid singularities at the inner boundary of the disk. We see from upper left panel a difference in the half thickness of the accretion disk for different eccentricities $e = 10^{-4}$ (black dashed-dotted curve), $e = 0.2$ (red dashed curve) and $e = 0.8345$ (solid blue curve). Higher eccentricity $e$ of the star corresponds to lower disk thickness at a particular radial distance from the center of MS. The half-thickness $z_0(r)$ of accretion disk increases rapidly at smaller radii but at large radii the disk thickness is almost constant and reduces to standard $\alpha$-disk. 
\begin{figure}
\centering
\includegraphics[width=1\columnwidth]{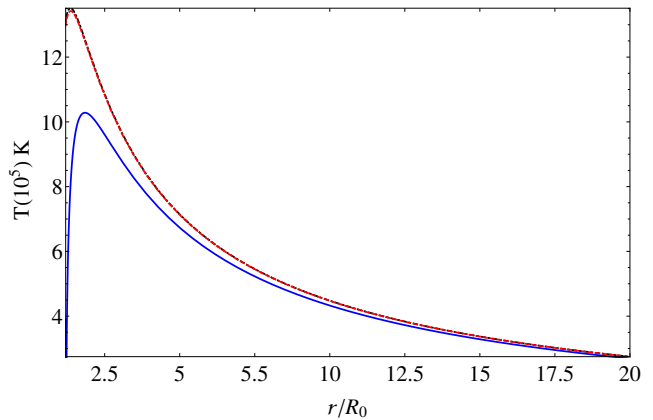}
\caption{\small{Plot shows the variation of photo-sphere temperature (Eq.~\ref{temp}) for eccentricities $e = 10^{-4}$ (black dotted dashed curve), $e = 0.2$ (red dashed curve) and $e = 0.8345$ (solid blue curve). The mass accretion rate is fixed at $\dot{M} = 10^{-4}\dot{M}_{Edd}$.}}
\label{fig4}
\end{figure}
\begin{figure}
\centering
\includegraphics[width=1\columnwidth]{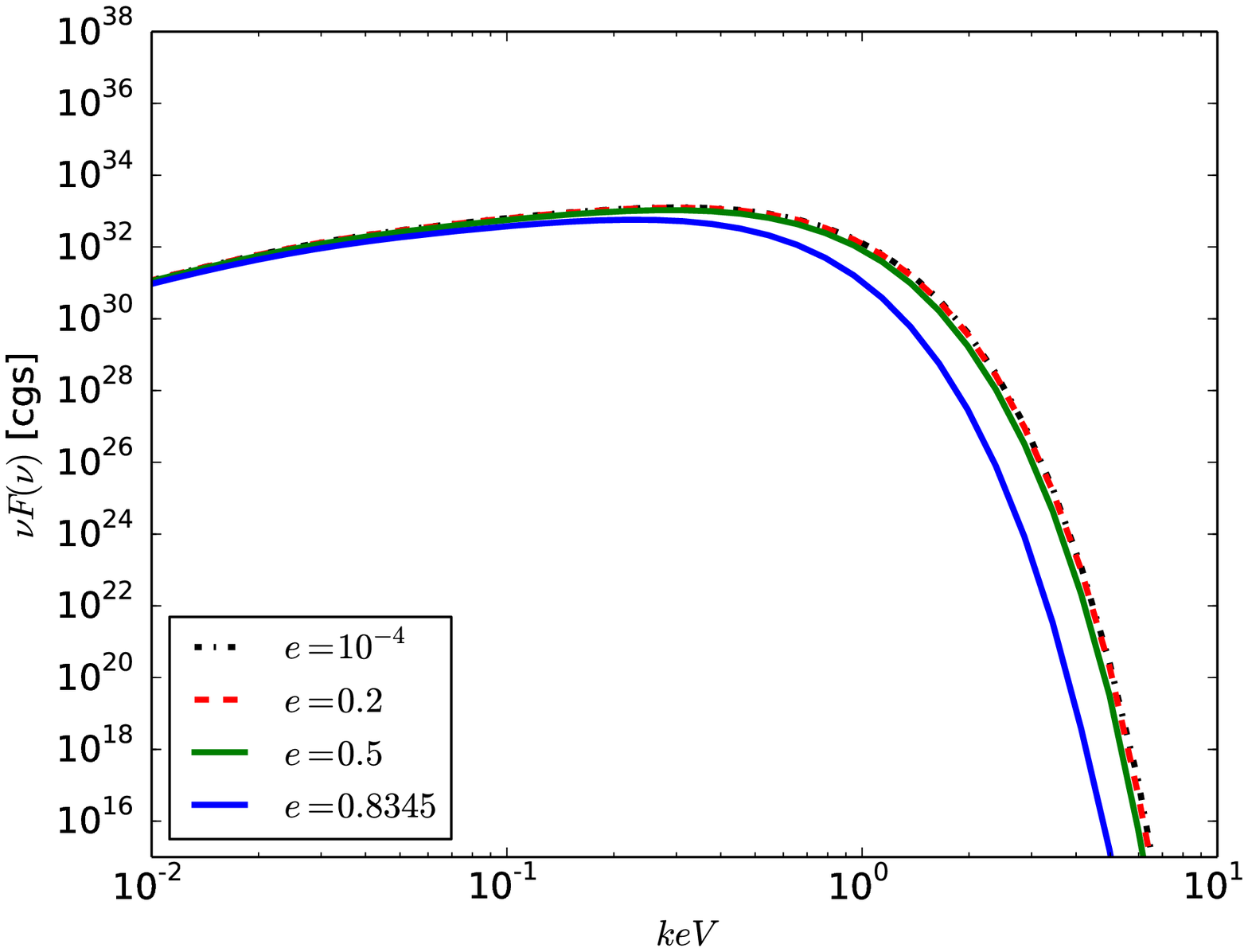}
\caption{\small{Logarithmic plot shows the emitted spectra from the accretion disk with mass accretion rate, $\dot{M} = 10^{-4}\dot{M}_{Edd}$. Four different eccentricities, $e = 10^{-4}$(black dotted-dashed) $e = 0.2$ (red dashed), $e = 0.5$ (solid green) and $e = 0.8345$ (solid blue) are shown in the plot.}}
\label{fig5}
\end{figure}
\begin{figure}
\centering
\includegraphics[width=1\columnwidth]{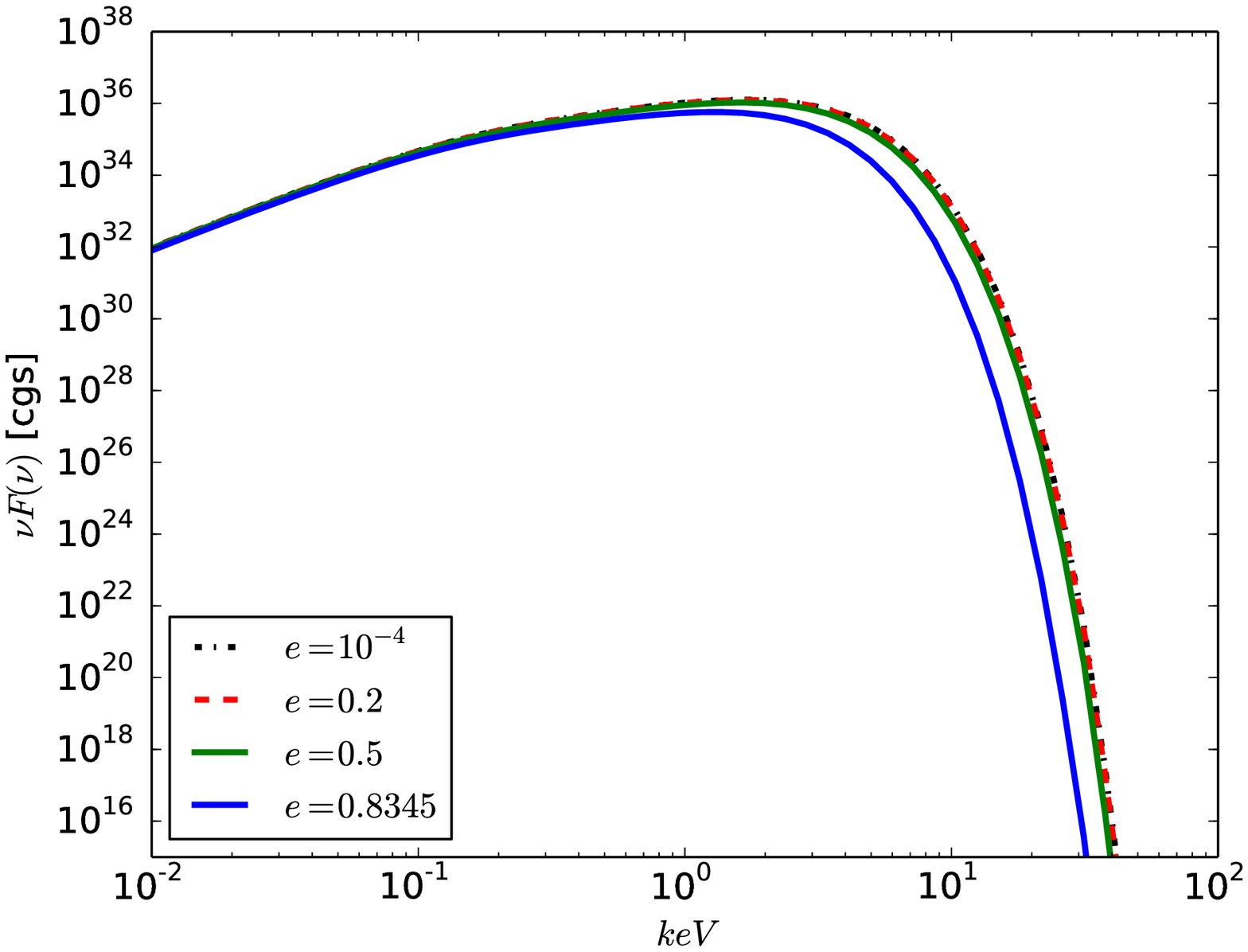}
\caption{\small{Logarithmic plot shows the emitted spectra from the accretion disk with mass accretion rate, $\dot{M} = 0.5\dot{M}_{Edd}$. Four different eccentricities, $e = 10^{-4}$ (black dotted-dashed) $e = 0.2$ (red dashed), $e = 0.5$ (solid green) and $e = 0.8345$ (solid blue) are shown in the plot.}}
\label{fig6}
\end{figure}
The upper right panel shows the logarithmic plot of the surface density distribution in the inner region of the accretion disk. We see that for $\dot{M} = 0.8\dot{M}_{Edd}$ mass accretion rate for higher eccentricities the surface density is higher than for the lower eccentricities. In case of low mass accretion rate , $\dot{M} = 10^{-4}\dot{M}_{Edd}$, the higher eccentricity corresponds to low surface density. The surface density $\Sigma(r)$ also increases with radial distance in the inner region of the accretion disk for $\dot{M} = 0.8\dot{M}_{Edd}$ mass accretion rate but for low mass accretion rate, $\dot{M} = 10^{-4}\dot{M}_{Edd}$, disk is gas pressure dominated. In gas pressure dominated region surface-density of the accretion disk decreases with radial distance. The lower left panel shows the radial variation of central temperature $T(r)$ in the accretion disk. In case of higher eccentricity the central temperature is lower. The lower right panel shows the radial velocity profile $v_r(r)$ in the accretion disk.
\subsection{Disk thermodynamics and Spectra}
The most important parameter for observational interest is emitted spectra from the accretion disk. Emitted spectra also corresponds to the size of accretion disk. We computed spectra using surface temperature of accretion disk. We chose a fixed outer radius of the accretion disk to see the behavior of the emitted spectra with change in eccentricity. The accretion disk we assumed here is optically thick in the $z$-direction therefore we can assume that each element of the disk emits as black body with surface temperature $T_s(r)$. Using angular velocity from Eq.~\ref{omega} and equating the dissipation rate per unit area to the black body flux we computed temperature of the accretion disk. Using the calculated temperature we can calculate intensity and with intensity, flux of emitted spectra from the accretion disk.
\begin{equation}
T_s(r) = \left[\frac{\dot{M}r\Omega T_1(r)}{4\pi\sigma}\frac{d\Omega}{dr}\right]^{1/4}
\label{temp}
\end{equation}
where $T_1(r)$ is defined as
\begin{equation}
T_1(r) = 1 - \frac{\Omega(a)a^2}{\Omega(r)r^2}
\label{T1}
\end{equation}
Now approximating the disk emitted spectra with black body we have
\begin{equation}
I(\nu) = B_{\nu}[T_s(r)] = \frac{2h\nu^3}{c^2(e^{h\nu/kT_s(r)} -1)}
\end{equation}
using Eq.~\ref{T1} we computed flux emitted from accretion disk by integration over the whole disk.
\begin{equation}
F(\nu) = 2\pi\int^{100R_0}_{a(e)}I(\nu)rdr
\label{flux}
\end{equation}
the integration in Eq.~\ref{flux} gives emitted spectra from the disk. Fig.~\ref{fig4} shows the variation of photo-sphere temperature. We see a significant change in photo-sphere temperature $T_s(r)$ unlikely central disk temperature, $T(r)$ (Eq.~\ref{midplane}). We also see from Fig.~\ref{fig4} that at large radii surface temperature $T_s(r)$ for different eccentricities correspond to same value. This is so because orbital frequency (Eq.~\ref{omega}) reduces to its Keplerian value at large radial distance. This corresponds to photo sphere temperature for accretion disk around spherical object. This can also be verified from Eq.~\ref{temp} by substituting orbital frequency as Keplerian orbital frequency for potential due to spherically symmetric central object. The central temperature and surface temperature has same behavior on eccentricity change, higher the eccentricity lower the temperature (Fig.~\ref{fig2}, Fig.~\ref{fig4}). We considered two different mass accretion rates in units of $\dot{M}_{Edd}\approx 10^{17}\textrm{g}$ $\textrm{s}^{-1}$ for chosen density and radius of MS. Fig.~\ref{fig5} and Fig.~\ref{fig6} shows the logarithmic variation of emitted spectra from the accretion disk for mass accretion rate $\dot{M} = 10^{-4}\dot{M}_{Edd}$ and $\dot{M} = 0.5\dot{M}_{Edd}$ with MS eccentricities $e = 10^{-4}$ (black dashed curve) $e = 0.2$ (red dashed curve), $e = 0.5$ (solid green curve) and $e = 0.8345$ (solid blue curve). The high frequency emission occurs at inner region of the disk where Maclaurin spheroid potential dominates. This gives small difference in high frequency region for emitted spectra of accretion disk. In our model we assumed that disk terminates on the surface of MS which has solid surface. This can cause formation of boundary layer region \citep{1995ApJ...442..337P, 1996ApJ...473..422P, 1998MNRAS.297..570T}. At this region the angular velocity of the accreting matter which is described by Eq.~\ref{omega} will change to angular spin rate of MS. If the angular spin rate of the MS, $\Omega_* < \Omega(R_0)$ it will form boundary layer region. In case of low eccentricity MS the angular velocity of matter will be significantly large as compare to angular spin of MS. This can also significantly affect the emitted spectra from the inner region of the accretion disk. In this paper we are only interested for spectra emitted from the disk due to dissipation in the disk only. The detailed discussion of boundary layer region will be addressed in the future work.
\begin{figure*}
\centering
\includegraphics[width=2\columnwidth]{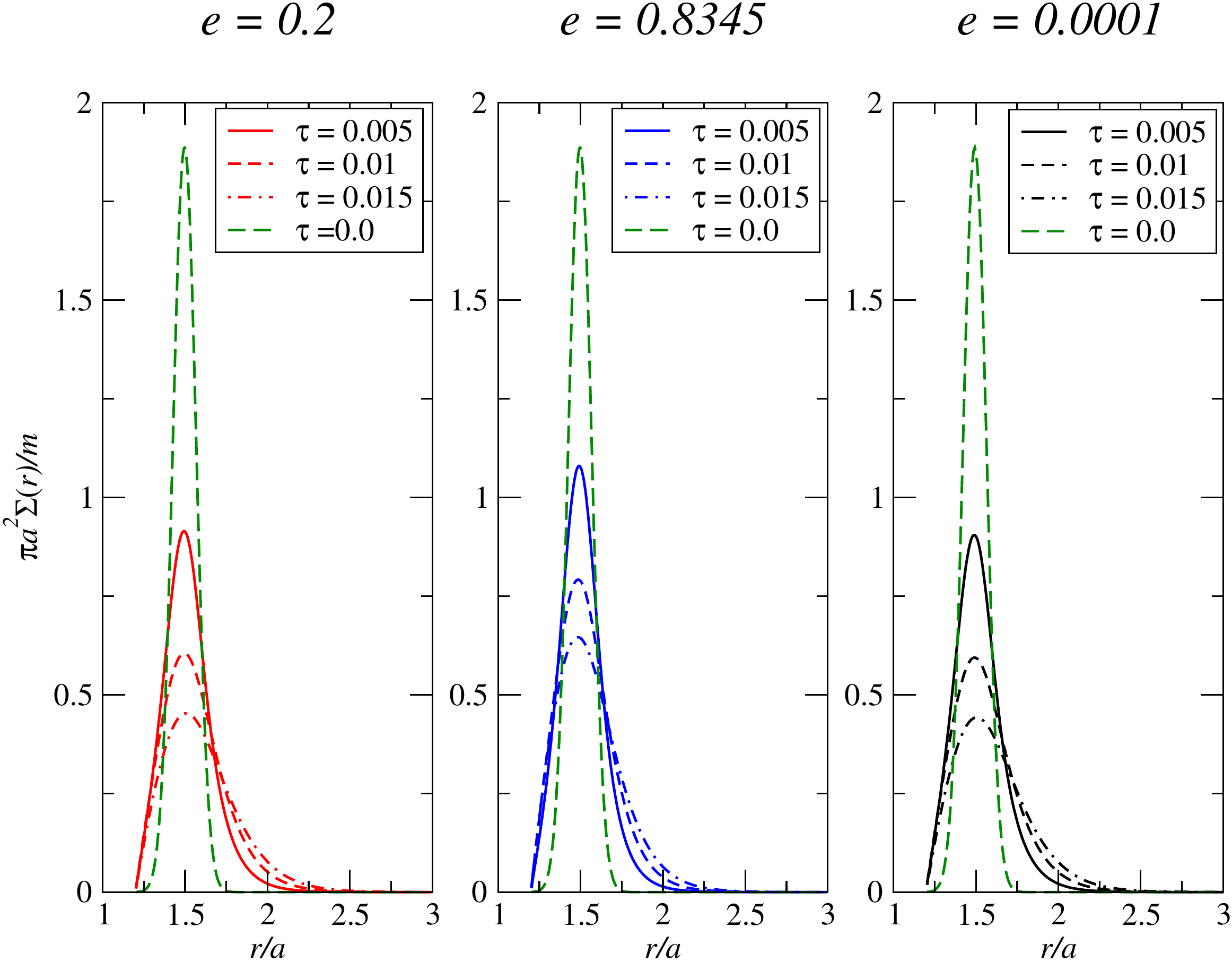}
\caption{\small{Time evolution of the ring of matter at a radial distance of $r = 1.5 a$. The vertical axis shows the surface density scaled with initial surface density of ring of matter with mass $m$. The horizontal axis corresponds to radial distance form the center of star. The leftmost panel shows the time evolution of the surface density $\Sigma$ for $e = 0.2$. The initial density Gaussian profile is shown by green long dashed curve in all three panels. In the left panel $\tau = 0.005$ (solid red curve), $\tau = 0.01$ (dashed red curve) and $\tau = 0.015$ (red dashed curve) are showing the time evolution of the ring. The middle panel shows the time evolution for $e = 0.8345$. The middle panel also shows the evolution for $\tau = 0.005$ (solid blue curve), $\tau = 0.01$ (dashed blue curve) and $\tau = 0.015$ (blue dashed curve are shown. The right most panel shows the time evolution for $e = 0.0001$ with same curve types for same evolution time of $\tau = 0.005$, $e = 0.01$ and $e = 0.015$ with black color.}}
\label{fig7}
\end{figure*}
\begin{figure}
\centering
\includegraphics[width=1\columnwidth]{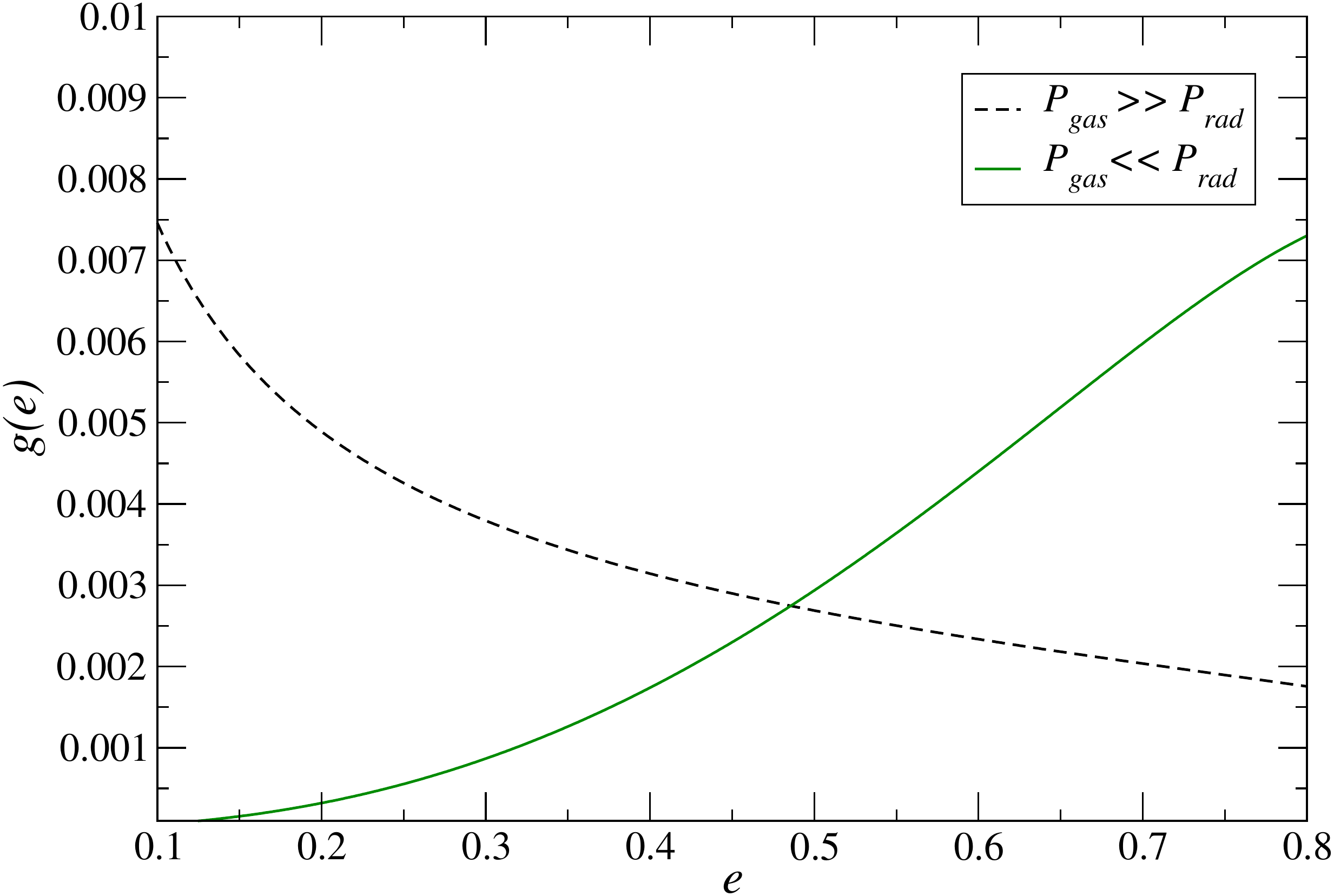}
\caption{\small{Plot shows eccentricity dependence of surface density, $\Sigma(r)$, at $r_0 = 5.0a$ in the radiation and gas pressure dominated results respectively. The $x$-axis shows the eccentricity, $e$ of the MS and $y$-axis shows the function of eccentricity corresponding to dependence of surface density, $\Sigma(r)$, on the eccentricity of MS.}}
\label{eccent}
\end{figure}
\begin{figure}
\centering
\includegraphics[width=1\columnwidth]{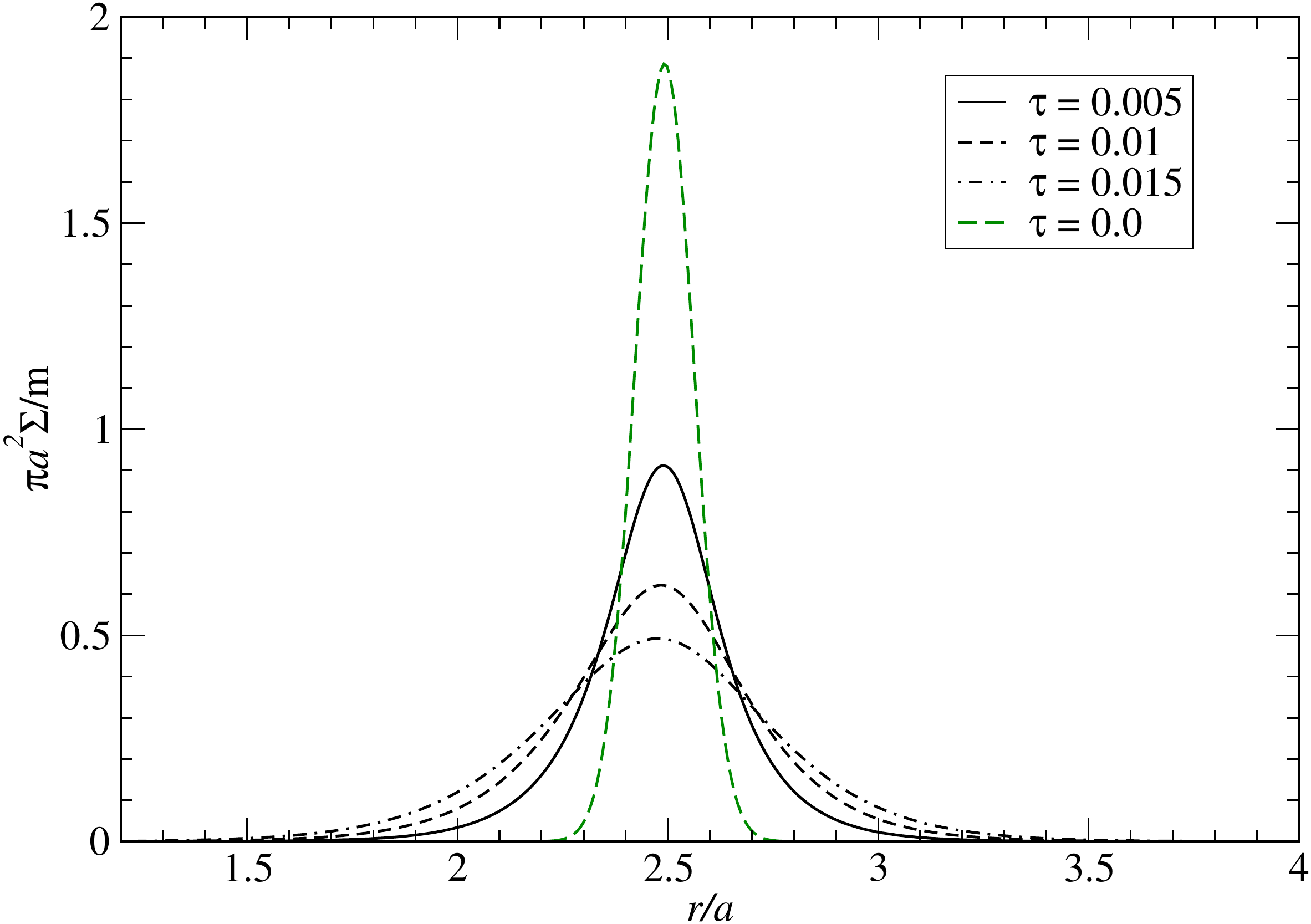}
\caption{\small{Plot shows the limiting case of $e = 10^{-4}$ for an initial ring of matter at $r = 2.5a$. Green long dashed curve corresponds for the initial Gaussian ring of matter. Solid black curve corresponds for viscous time $\tau = 0.005$, black dashed curve corresponds to $\tau = 0.001$ and black dotted-dashed curve corresponds to $\tau = 0.015$. The vertical scale is scaled with initial surface density of the Gaussian ring of mass $m$. The radial distance is scaled with semi-major axis of the MS. The evolution time $\tau$ is expressed in units of viscous time.}}
\label{fig8}
\end{figure}
\subsection{Evolution of surface density}
We used the physical model of non-stationary accretion disk described in \S2.3 to study the time evolution of accretion disk. We numerically integrated Eq.~\ref{diffusion} and Eq.~\ref{velocity} by using Crank-Nicolson algorithm. We assumed an initial Gaussian density distribution of matter at a radial distance of $r = 1.5a$ as the initial condition to solve the diffusion equation (Eq.~\ref{diffusion}). In all the results of non-stationary disk the time is in viscous time scale, $t_{visc} = a^2/\nu$, $a$ being the semi-major axis of the star. In our model we are interested for constant viscosity prescription therefore the kinematic viscosity coefficient $\nu = 0.01$ (in code units) is constant throughout our numerical computation. Similar to our analytical model we covered the range of eccentricity from lower value $e = 0.0001$ to critical limit $e = 0.8345$ with an intermediate value of $e = 0.2$. 

In Fig.~\ref{fig7} we plotted the time evolution of surface density for the ring of matter at $r = 1.5a$ for $e = 0.0001$, $e = 0.2$ and $e = 0.8345$. The vertical axis shows the surface density $\Sigma(r)$ scaled with initial surface density. The horizontal axis corresponds to radial distance from the center of star scaled with semi-major axis $a$ of the MS. The leftmost panel shows the time evolution of accretion disk with $e = 0.2$ for viscous time $\tau = 0.005$ (solid red curve), $\tau = 0.01$ (dashed red curve) and $\tau = 0.015$ (dotted-dashed red curve). The green long dashed curve in all the panels shows the initial Gaussian density distribution at $\tau = 0.0$. The middle plot shows the time evolution in case of $e = 0.8345$ for time steps of $\tau = 0.005$ (solid blue curve), $e = 0.01$ (blue dashed curve) and $e = 0.015$ (blue dotted dashed curve). We showed from left and middle panel that if the eccentricity is high the matter will diffuse rather slowly as compared to lower eccentricities (inverse law). The rightmost panel shows the limiting case $e = 0.0001$ to compare the results from an accretion disk evolution in case of spherically symmetric objects. Similar to left and middle panel rightmost panel shows the time evolution for viscous time $\tau = 0.005$ (solid black curve), $\tau = 0.01$ (dashed black curve) and $\tau = 0.015$ (black dotted-dashed curve). 

In Fig.~\ref{fig8} we also showed the behavior of the Gaussian ring of matter at $r = 2.5a$ for an eccentricity of $e = 0.0001$. The goal to present this figure is to avoid the confusion from Fig.~\ref{fig7}. The matter in Fig.~\ref{fig8} spread almost symmetrically around $r = 2.5a$. We also tested our numerical code by reducing the code parameters to the limiting case of existing results of spherically symmetric potential. In the appendix various code parameters are defined in which we chose parameter $D$ (Eq.~\ref{D}) which corresponds to the diffusion of matter. Fig.~\ref{fig9} shows the variation of $D$ with radial distance for eccentricities $e = 0.0001$ (black dotted-dashed curve), $e = 0.2$ (red dashed curve) and $e = 0.8345$ (solid blue curve). The limiting value in case of Keplerian angular velocity or $e = 0.0$ is $D = -3.0$. We see from Fig.~\ref{fig9} that as we decrease the eccentricity, parameter $D$ is converging to $D = -3.0$. Also for larger radial distance the parameter $D$ converges to the limit of spherically symmetric potential ($D = -3.0$).
\section{Discussion}
\subsection{Steady thin disk}
We found that our approximation of potential due to Maclaurin spheroid (MS) cause changes in the steady state thin disk parameters, half thickness $z_0(r)$, surface density $\Sigma(r)$ and central disk temperature $T(r)$. The disk will be radiation pressure dominated in the very inner regions for a mass accretion rate, $\dot{M} = 0.8\dot{M}_{Edd}$. We also calculated disk quantitites for mass accretion rate, $\dot{M} = 10^{-4}\dot{M}_{Edd}$  where the disk is gas pressure dominated (opacity is due to electron scattering). Further depending on mass accretion rate disk quantities will be either governed by radiation pressure dominated calculations (Eq.~\ref{z} to Eq.~\ref{vel}) or gas pressure dominated calculations (Eq.~\ref{gasd} to Eq.~\ref{gasv}). 

In both scenarios, a change in eccentricity from $e = 10^{-4}$ to $e = 0.2$ gives almost no change in the calculated steady state thin disk parameters. We also see from Fig.~\ref{fig2} and Fig.~\ref{fig3} that an increase in eccentricity has same behavior as decrease in mass accretion rate. For example from Eq.~\ref{sigma} $\Sigma(r)\propto\dot{M}^{-1}$ thus decrease in mass accretion rate would increase the surface density, similarly increase in surface density is seen with increase in eccentricity (top right panel Fig.~\ref{fig2}). However in gas pressure dominated disk surface density decreases with eccentricity (see Fig.~\ref{fig3}) and verifies the fact that $\Sigma\propto\dot{M}^{3/5}$ (see Eq.~\ref{gasd}). Thus radiation pressure and gas pressure dominated disk show opposite behavior with regards to dependence on eccentricity not only in $\Sigma(r)$ but also in $v_{r}(r)$ and $z_0(r)$ (see Fig.~\ref{fig2} and Fig.~\ref{fig3}). This difference can also be quantitatively investigated with the in-depth study of dependence of these quantities on the orbital frequency, $\Omega(r)$, for MS potential decreases with eccentricity, $e$ \citep{2013MNRAS.434.2825K}. In general,  the surface density at a particular disk radius $r_0$ is given by,
\begin{equation}
\Sigma(e, r_0) = \Lambda \textrm{g}(e, r_0)
\end{equation}
where, $\Lambda$ is a numerical constant that also incorporates the steady mass accretion rate. 
The function $\textrm{g}(e, r_0)$ represents the eccentricity dependence and will be different depending on whether disk is gas pressure dominated or radiation pressure dominated.

The variation of the function $\textrm{g}(e, r_0)$ with eccentricity for $r_0 = 5.0a$  is shown in Fig.~\ref{eccent} for radiation (\textit{green line}) and gas (\textit{black dashed}) pressure dominated disks. The curve evidently shows the opposite behavior in the eccentricity dependence for the function $\textrm{g}(e, r_0)$ (or $\Sigma(e, r_0)$) in agreement to variations seen in Fig.~\ref{fig2} and Fig.~\ref{fig3}. In a similar manner, this analysis can be extended to the radial velocity and half thickness of the disk. However the disk central temperature show similar trends with dependence on eccentricity for radiation and gas pressure dominated disk. In case of gas pressure dominated disk $T(r)\propto\dot{M}^{2/5}$, since increasing eccentricity implies to decrease in mass accretion rate, $T(r)$ decreases with eccentricity seen in Fig.~\ref{fig3}. Whereas $T(r)$ in radiation pressure dominated disk is independent of $\dot{M}$ and the net dependence on eccentricity comes from product of energy flux radiated from the disk and surface density. This product cancels the increasing trend of surface density with eccentricity resulting in lower disk central temperature with higher eccentricity similar to gas pressure dominated disk.
\begin{figure}
\centering
\includegraphics[width=1\columnwidth]{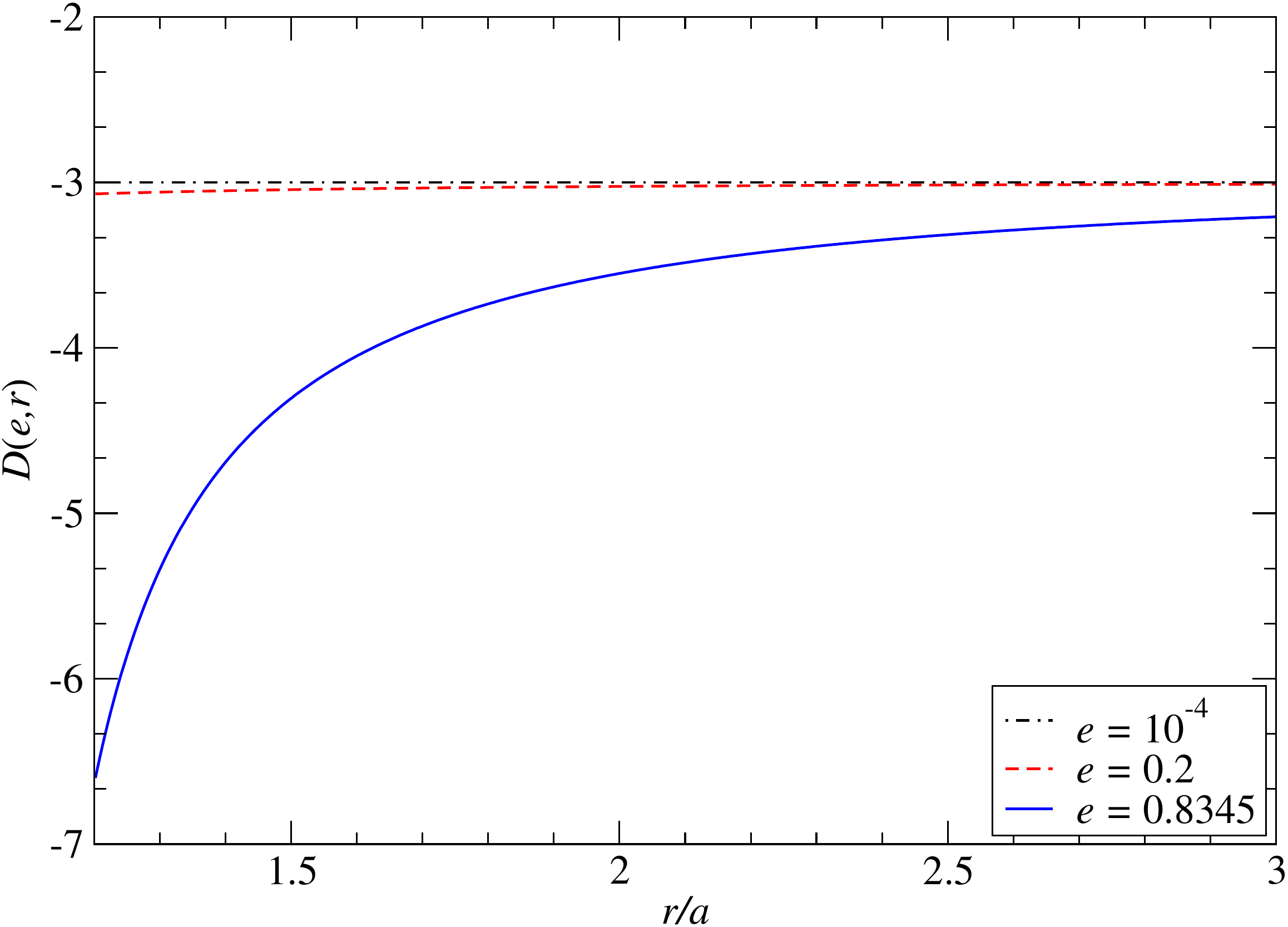}
\caption{\small{Plot shows the test of our code in the limiting case when $e \rightarrow 0$ for spherically symmetric object. The vertical axis shows the variation of the parameter $D$ defined in Eq.~\ref{D}. The horizontal axis shows the radial distance form the center of MS. Three different eccentricities $e = 10^{-4}$ (black dotted-dashed curve), $e = 0.2$ (red dashed curve) and upper limit in our model $e = 0.8345$ are plotted.}}
\label{fig9}
\end{figure}
\subsection{Dependence of Spectra on eccentricity}
\label{dspectra}
We computed the disk spectra for different values of eccentricity $e = 10^{-4}$, $e = 0.2$, $e = 0.5$ and $e = 0.8345$ of MS. We know from the existing results that a change in the disk area changes the emitted spectra. In this paper the only parameter we changed is the eccentricity and since a change in the eccentricity of the MS is changing the semi-major axis of the star. The assumption in our model that thin disk terminates at the surface of the MS changes the inner radius of the accretion disk and in result gives minor changes to the emitted spectra from the inner region of accretion disk. Quantitatively the inner radius of disk for $e = 10^{-4}$ is $a = 10^6 \textrm{cm}$ (as disk terminates on the surface of MS) for $e = 0.2$ it is $a = 1.007\times 10^6$ and for $e = 0.8345$ it is $a = 1.219\times 10^6 \textrm{cm}$. This gives minor change in disk area and so emitted spectra of the disk. 

The photo sphere temperature also changes with change in eccentricity (Fig.~\ref{fig4}) which affect the emitted spectra at high frequency region (Fig.~\ref{fig5} and Fig.~\ref{fig6} ). In Eq.~\ref{temp} increasing the mass accretion rate $\dot{M}$, will increase the surface temperature. Increase in surface temperature will cause the peak of spectra to move towards higher frequency. An increase in eccentricity of MS will decrease the surface temperature $T_s(r)$ and so the peak of the spectra will move towards lower frequency (Fig.~\ref{fig5} and Fig.~\ref{fig6}). We also considered two different mass accretion rates one mass accretion rate $\dot{M} = 10^{-4}\dot{M}_{Edd}$ (Fig.~\ref{fig5}) and $\dot{M} = 0.5\dot{M}_{Edd}$ (Fig.~\ref{fig6}). In case of $\dot{M} = 10^{-4}\dot{M}_{Edd}$, mass accretion rate we found that the peak of the spectra emitted from the disk lies at approximately $1.0$ $\textrm{keV}$ in X-ray band and for $\dot{M} = 0.5\dot{M}_{Edd}$ the peak lies at approximately $5.0$ $\textrm{keV}$ which is also in X-ray band. We can also say qualitatively from above discussion that an increase in eccentricity of MS has same behavior as decrease in mass accretion rate. The accretion luminosity for $\dot{M} = 0.5\dot{M}_{Edd}$ is of the order of $L_{acc} = 0.1L_{Edd}$. The disk luminosity may have small changes due to change in eccentricity of the MS. The emitted spectra may help to determine the eccentricity and so period of the MS. The period of the star can be correlated to eccentricity of MS using following expression \citep{1990ApJ...359..444F} 
\begin{equation}
P \lesssim 0.66\textrm{ms}\left(\frac{M_*}{1.4M_{\odot}}\right)^{-1/2}\left(\frac{R_0(1 - e^2)^{-1/6}}{10 \textrm{km}}\right)^{3/2}
\label{spin}
\end{equation}
where $R_0$ is the semi major axis of MS for $e = 0.0$. If we assume that our results will hold for neutron stars. This model may also help to constrain the equation of state and density distribution in the neutron stars \citep{1998ApJ...509L..37K, 1999A&A...352L.116S}. The determined period and equation of state or density distribution may help to discriminate between neutron star, quark star and white dwarf. The estimation of eccentricity of MS with the help of disk spectra may also help to understand whether ISCO lies above the surface of star or not.
\subsection{Non-stationary disk}
The results of non-stationary disk are also dependent on eccentricity of MS. This is indeed due to different choice of potential in our model. We found that if the eccentricity of the central object is lower the viscous evolution of the accretion disk will be more rapid as compared to high eccentricity. The choice of location of initial Gaussian distribution of matter is also important in our numerical model. We kept initial distribution of matter at $r = 1.5a$, which is very close to the MS. If we start at large radial distance as we can see from Fig.~\ref{fig9}, the effect of eccentricity change will not be significant. From this result we can also explain that a spin-up or spin-down of the rapidly rotating MS can change the viscous evolution of the accretion disk. If the MS spin-up eccentricity will increase and in result the diffusion will be slower and if spin-down i.e. decrease in eccentricity it will cause rapid evolution of the accretion disk. We also see from Fig.~\ref{fig9} that a change in eccentricity changes the diffusion coefficient. 

We can also study the variability phenomenon in the accretion disk \citep{1976MNRAS.175..613S}. The change in diffusion term $D$ can cause the variability in the accretion disk. In this way a change in eccentricity may either slow down the diffusion or enhance the diffusion. In non-stationary disk also we see that an increase in eccentricity of MS has same behavior as decreasing the mass accretion rate. Although time scale to change the eccentricity of the MS is much larger than the dynamical time scale of the disk but it may help to understand variability phenomenon occurring in the accretion disk.
\section{Conclusions}
\begin{itemize}
\item[1.] We used Maclaurin spheroid potential to study geometrically thin accretion disk. Semi-analytic work is done for the steady state disk and numerical work has been carried out to study non-stationary behavior of the disk.
\item[2.] We found that qualitatively a change in eccentricity plays a same role as decrease in mass accretion rate. We confirmed this behavior by computing emitted spectra from the accretion disk. This is also clear from the various disk parameters presented in Fig.~\ref{fig3}.
\item[3.] We also found in our thin disk analytical calculations that dependence of disk parameters like thickness, surface density and radial velocity on eccentricity of MS is opposite for radiation and gas pressure dominated disk (opacity due to electron scattering). The disk central temperature has the same dependence on eccentricity, higher the eccentricity lower the disk central temperature.
\item[4.] We numerically solved the non-stationary accretion disk and found that an increase in luminosity can slow down the diffusion of the matter.
\item[5.] We also suggested astrophysical applications of our results. The emitted spectra in our results can help to estimate the eccentricity thus period of neutron stars or quark stars. Period of star may help to understand the equation of state of neutron star or quark star.
\end{itemize}
\appendix
\section{}
\subsection{Steady-state disk}
\label{a1}
\subsubsection{Gas pressure dominated region, free-free absorption}
In this region similar to gas pressure dominated region $P_g >> P_r$ but the opacity is determined by free-free absorption. Sound speed is defined as $v^2_s = kT/m_p$.
\begin{equation}
\Sigma (r) = \left(\frac{m_p}{(2\pi\alpha k)^{4/5}}\right)\left(\frac{32\pi r bc \dot{M}^7\left(\Omega r^2 - \Omega(a) a^2\right)}{3\Omega\frac{d\Omega}{dr}}\right)^{1/10}
\end{equation}
\begin{equation}
T(r) = \left(\frac{3Q \Sigma(r)^{1/2}\Omega}{2 m_p^{1/2}}\right)^{1/8}
\end{equation}
\begin{equation}
z_0(r) = \sqrt{\frac{k T(r)}{m_p\Omega^2}}
\end{equation}
\begin{equation}
n(r) = \frac{\Sigma(r)}{2m_p z_0(r)}
\end{equation}
\begin{equation}
\tau = \sigma_{ff}\Sigma(r)
\end{equation}
\begin{equation}
v_r = \frac{\dot{M}}{2\pi\Sigma(r) r}
\end{equation}
where 
\begin{equation}
Q = -\frac{\dot{M}\left(\Omega r^2 - \Omega(a)a^2\right)}{4\pi r}\frac{d\Omega}{dr}.
\end{equation}
is the energy flux radiated from unit surface of disk in unit time and rest of the parameters defined above have the same notation as in the radiation pressure dominated region of the accretion disk.
\subsection{Non-stationary accretion disk}
The Crank-Nicolson algorithm we used has been already described in very detail for solving a general form of advection-diffusion equation \citep{2010A&A...513A..79B}. Here we shall describe the different terms which we calculated for our model to fit with Eq.~\ref{advdiff}  
\begin{equation}
\frac{\partial\Sigma}{\partial t} + \frac{\partial}{\partial r}(\Sigma u) - \frac{\partial}{\partial r}\left[h D\frac{\partial}{\partial r}\left(\nu\frac{\Sigma}{h}\right)\right] = L\Sigma
\label{advdiff}
\end{equation}
where
\begin{equation}
D = \frac{r\Omega^\prime}{\left(2\Omega + r\Omega^\prime\right)}
\label{D}
\end{equation}
\begin{equation}
u = \frac{\nu\Omega^\prime}{\left(2\Omega + r\Omega^\prime\right)}
\end{equation}
\begin{equation}
h = 1/(r^3\Omega^\prime)
\end{equation}
\begin{equation}
L = -\nu r^3\Omega^\prime\left[\frac{3}{r^4 l_1} +\frac{3\Omega^\prime + r\Omega^{\prime\prime}}{r^3l_1^2}\right]
\end{equation}
where $\Omega^\prime$ is the derivative with respect to radial distance from center of MS and
\begin{equation}
l1 = (2\Omega + r\Omega^\prime)
\end{equation}
\label{lastpage}
\section*{Acknowledgments}
We thank to Wlodek Klu\'zniak for proposing this project. We are also specially thankful to editor and referee for important suggestions and comments. We also thank to F.H. Vincent and A. Manousakis for discussions. Research was supported by Polish NCN grant UMO-34 2011/01/B/ST9/05439 and 2013/08/A/ST9/00795.
\bibliographystyle{mn2e}
\bibliography{ref}
\end{document}